# Pathway-based feature selection algorithms identify genes discriminating patients with multiple sclerosis apart from controls

Lei Zhang[1,2], Linlin Wang[1], Pu Tian[1§], Suyan Tian[3§]

[1] School of Life Science, Jilin University, 2699 Qianjin Street, Changchun, Jilin, China, 130012
[2] Department of Neurology, The Second Hospital of Jilin University, 218 Ziqiang Street, Changchun, Jilin, China, 130041
[3] Division of Clinical Epidemiology, The First Hospital of Jilin University, 71Xinmin Street, Changchun, Jilin, China, 130021

[§]Corresponding authors

Email addresses:

    ST: stian@rockefeller.edu
    PT: tianpu@jlu.edu.cn

# Abstract

**Introduction**

The focus of analyzing data from microarray experiments and extracting biological insight from such data has experienced a shift from identification of individual genes in association with a phenotype to that of biological pathways or gene sets. Meanwhile, feature selection algorithm becomes imperative to cope with the high dimensional nature of many modeling tasks in bioinformatics. Many feature selection algorithms use information contained within a gene set as a biological priori, and select relevant features by incorporating such information. Thus, an integration of gene set analysis with feature selection is highly desired. Significance analysis of microarray to gene-set reduction analysis (SAM-GSR) algorithm is a novel direction of gene set analysis, aiming at further reduction of gene set into a core subset. Here, we explore the feature selection trait possessed by SAM-GSR and then modify SAM-GSR specifically to better fulfill this role.

**Results and Conclusions**

Training on a multiple sclerosis (MS) microarray data using both SAM-GSR and our modification of SAM-GSR, excellent discriminative performance on an independent test set was achieved. To conclude, absorbing biological information from a gene set may be helpful for classification and feature selection.

**Discussion**

Given the fact the complete pathway information is far from completeness, a statistical method capable of constructing biologically meaningful gene networks is in demand. The basic requirement is that interplay among genes must be taken into account.

# Introduction

With the development of several major databases which organize different types of biological pathway or gene set information, e.g., **Kyoto Encyclopedia of Gene and Genomes** (KEGG) [1] and **Gene Ontology** (GO) [2], the coordinated effect of a pathway or gene set as a whole has been explored. Subsequently, many pathway or gene-set based methods have been proposed [3–11]. Meanwhile, the focus of analyzing data from microarray experiments and extracting biological insight from such data have experienced a shift from identification of individual genes in association with a phenotype to that of biological pathways or gene sets.

A gene set analysis requires extra consideration than individual gene analysis, namely, the incorporation of the whole gene set into an association measure [12]. A gene set analysis can be stratified into two major categories based on the formulation of null hypothesis, namely, a 'self-contained' test and a 'competitive' test [13]. While the former only considered the expression values of genes within a gene set, the latter required some comparison between the gene expression values of gene set being tested and those of the genes outside. Many researchers give preference to a 'self-contained' test [14]. However, the Gene Set Enrichment Analysis (GSEA) [5], a more competitive-like gene set test, gains the top popularity. Despite that GSEA has its own advantages, it has been widely criticized for suffering many drawbacks [10,12,15] even after major modifications had been specifically made to address some of those. In this article, without causing confusion the phrases "gene set" and "pathway" are used interchangeably.

Nowadays, feature selection algorithm becomes imperative to cope with the high dimensional nature of many modeling tasks in bioinformatics [16]. Among a variety of feature selection algorithms, many use information contained within a gene set as a biological priori and select relevant features by incorporating such information. An important finding is the incorporation of such gene set information can potentially increase the predictive power of a classifier and identify more biological meaningful features [8,17,18]. For instance, Ma et al [19] proposed a supervised group Lasso method which can divide genes into clusters by incorporating the gene set information. Briefly, this method consists of two steps. First, identification of relevant genes within each cluster using Lasso method was conducted. Then the effort was devoted to select relevant clusters using the group Lasso. Of note, the authors mentioned they defined cluster structure based on statistical measurements such as a K-mean method, given gene set information is only partially available or even not available for the genes under consideration. Thus the clusters constructed by this means are mutually exclusive and no genes are excluded due to lack of biological information in the annotated databases. Nevertheless, in reality it is routine to have a single gene involving in many gene sets or pathways. Another example is the one suggested by [20]. In this algorithm, a pseudo-gene taking the average value of all genes within a gene set was created to represent the whole gene subset, and then the downstream analysis was conducted using those pseudo-genes. Such a method leads to results with poor biological interpretation, and is in vain for selecting relevant genes.

A more relevant direction of gene set analysis was proposed by [15], which aimed at further reduction of gene set into a core subset. The authors claimed that reducing a significant gene set to core subset is an essential step towards understanding biological mechanisms underlying the gene-set association with the phenotype of interest. The reasons they gave to support this statement include 1) a smaller set of genes is easier to understand and facilitate biological insight into disease processes, and 2) reduction to the most predictive genes allows for targeted therapies and intervention strategies, and 3) such reduction facilitates a downsize of platform from a high-throughput microarray technology to cheaper and quicker methods such as real-time PCR, thus increases the applicability for diagnostic purposes in a clinical setting. The proposed method was named as significance analysis of microarray to gene set reduction analysis (SAM-GSR). Obviously, those issues are also targeted by a feature selection algorithm, which motivates us to conduct feature selection based on the selected gene sets, using SAM-GSR algorithm.

Multiple sclerosis (MS) is the most prevalent demyelinating disease and the principal cause of neurological disability in young adults [21]. Diagnosis of MS can only be confirmed currently using invasive and expensive tests such as magnetic resonance imaging (MRI). Therefore, researchers have beginning to resort to other technologies for an easier and cheaper diagnosis of MS. Among them, microarray technology is the one being extensively explored [22–24] even though compared to its popularity in cancer research, the number of microarray experiments on MS is limited and the sample sizes of those studies are predominately small [25]. A combination of an integrated microarray data set and a suitable computational algorithm to classify MS samples is highly desirable. Fortunately, as a part of the recently-launched systems biology verification (sbv) Industrial Methodology for Process Verification in Research (IMPROVER) Challenge [26], MS diagnosis sub-challenge targeted specifically on this issue, using gene expression data obtained from microarray experiments. Among the challenge participants who ranked high in this sub-challenge, two pieces of work were relevant to gene-set based analysis. First, Lauria [27] used Cytoscape [28] to construct two separate clusters/networks to discriminate MS samples from controls. However, the modeling parsimony was not a concern in this method. Therefore, the results might be not applicable in the clinical setting. In contrast, Zhao et al [29] implemented the method by Chen et al. [20] and generated one pseudo-gene for each pathway by averaging expression values of all genes in that pathway. Then a logistic regression with elastic net regularization with the resulting pseudo-features was fitted. Nevertheless, it turned out such method was inferior to their final model in which the individual genes serving as covariates in a regularized logistic regression model.

In this paper, we explore the feature selection trait possessed by SAM-GSR and make specifically some modification on it to better fulfill this role. The results from our analyses indicate absorbing biological information from a gene set may be helpful to identify relevant genes with good discriminative performance.

## Methods and Materials

**Experimental data**
The training data includes chips from the experiment E-MTAB-69 stored in the ArrayExpress [30] repository (http://www.ebi.ac.uk/arrayexpress). All chips were hybridized on Affymetrix HGU133 Plus 2.0 chips. In this study, there were 26 patients with relapsing-remitting multiple sclerosis (RRMS) and 18 controls with neurological disorders of a non-inflammatory nature.

The test dataset is the one used in the IMPROVER MS sub-challenge, which is accessible to the challenge participants on the project website (http://www.sbvimprover.com). It was hybridized on Affymetrix HGU133 Plus 2.0, and there were 28 patients with RRMS and 32 normal controls.

The gene sets were downloaded from the **Molecular Signatures Database** (MSigDB) [5]. First, we considered c2 category. This category includes gene sets from curated pathways databases such as KEGG and those manually curated from the literature on gene expression. The current version (version 4.0) of MSigDB c2 category included 4,722 gene sets annotating on 11,844 unique genes. Similarly, the genes sets in c5 category were downloaded as well, including 1,454 gene sets annotated by GO terms.

**Pre-processing procedures**
Raw data of the training set was downloaded from the ArrayExpress repository, and expression values were obtained using the GCRMA algorithm [31] and normalization across samples was carried out using quantile normalization. The resulting expression values were on $\log_2$ scale. When there were multiple probe sets representing the same gene, the one with the largest fold change as indicated by the moderated t-tests was chosen. Then the resulting expression values of 19,851 unique genes were fed into downstream analysis. Raw data of the test set was downloaded from the sbv challenge website, and was pre-processed in the same way.

**Methods**

*SAM-GSR*

As mentioned in the **Introduction** section, SAM-GSR is an extension of SAM-GS with the objective of core subset identification. Briefly, the following statistic was defined for gene set *j* in SAM-GS,

$$SAM-GS_j = \sum_{i=1}^{|j|} d_i^2 \ , \quad d_i = (\bar{x}_d(i) - \bar{x}_c(i))/(s(i) + s_0)$$

where $d_i$ is SAM statistic [32] and calculated for each gene involving in a gene set, $\bar{x}_d(i)$ and $\bar{x}_c(i)$ are the sample averages of gene *i* respectively for diseased group and control group, s(i) is a pooled standard deviation and estimated by pooling samples over two groups, $s_0$ is a small positive constant used to offset the small variability in microarray measurements, and |j| is the size of gene set *j*. Thus, SAM-GSR statistic for a gene set is the $L_2$ norm of SAM statistics for all genes within the gene set.

Upon a significant gene set, where statistical significance is estimated using a permutation test by perturbing phenotype-labels for several hundred times, SAM-GSR gradually partitions the entire set S into two subsets: the reduced subset $R_k$ and the residual one $\bar{R}_k$ for k=1,…, |j|. Here, genes in gene set *j* are ordered decreasingly based on the magnitude of SAM statistic $d_i$. Let $c_k$ be the SAM-GS p-value of $\bar{R}_k$, the final $R_k$ corresponds to the least k so that $c_k$ is larger than a threshold. The choice for this threshold is very liberal and varies from one application to another. For more details on SAM-GSR, the original work [15] is referred.

*Our modifications to SAM-GSR*

In SAM-GSR, the significant level of a gene within a gene set is determined by its contribution to the SAM-GS statistic. It implies that if in a gene set $|d_i| > |d_j|$ for genes *i* and *j*, gene *j* is not be inside the reduce subset when gene *i* is not. However, when the goal is feature selection, the magnitude of individual SAM statistic is not the focus. Therefore, we propose to use a penalized machine learning method to perform feature selection automatically and classify samples simultaneously. Because support vector machine (SVM) [33] is one of widely used and extremely powerful supervised learning methods, especially suitable for the two-class classification tasks of microarray data [34], we propose to use a SVM classifier with a Smoothly Clipped Absolute Deviation (SCAD) [35,36] function as the penalty to do feature selection. The final tuning parameter in SVM-SCAD was selected using 5-fold cross validation. The extension of SAM-GSR for feature selection purpose is referred to as modified SAM-GSR herein.

Notably, because in SAM-GSR posterior probability of each sample is not computable, we use SVM to do classification and posterior probability calculations. The choice of SVM here is to keep consistent to analyses using modified SAM-GSR. Moreover, the cutoff values of $c_k$ in SAM-GSR are different among individual analysis. A grid of values (i.e., 0.05 to 0.5 with an increment of 0.05) is considered and the one achieving the best classification rate and the smallest size on training set is chosen. Figure 1 illustrates graphically on both SAM-GSR and modified SAM-GSR algorithms.

*Weighted co-expression networks*
Weighted co-expression network (http://www.genetics.ucla.edu/labs/horvath/CoexpressionNetwork) is one method to build mutually exclusive networks describing co-expression patterns among multiple genes. Because pathway information contained in the MSigDB database is incomplete or not available for many genes, those genes are consequently eliminated from the downstream analyses. To evaluate if such elimination causes any miss-out of potential relevant genes, we use the gene sets constructed by the weighted co-expression network algorithm to reanalyze the data, and make comparisons. For a brief description on the weighted co-expression network algorithm, the article by Ma et al. [8] is a good reference.

*Statistical Metrics*
Usually, using a single metric to evaluate an algorithm tends to produce bias, where an algorithm may be erroneously claimed to be superior if a metric favouring it is chosen [26]. Thus we use four metrics, namely, *Belief Confusion Metric* (BCM), *Area Under the Precision-Recall Curve* (AUPR), *Generalized Brier Score* (GBS), and error rate, to evaluate the performance of a classifier. For the detailed description on GBS, the following work [37,38] is referred. For BCM and AUPR, two metrics used by SBV challenge, their definition and interpretation are available in the SBV homepage (http://www.sbvimprover.com/sites/default/files/scoring_metrics.pdf). Briefly, BCM captures the average belief/confidence that a sample belongs to a class when indeed it belongs to this class, while AUPR summarizes the ability to correctly rank the samples known to be in a given class when sorted by the belief values decreasingly for that class, as summarized by [39].

**Statistical language and packages**
Statistical analysis was carried out in the R language version 3.0 (www.r-project.org), and R codes for SAM-GS and SAM-GSR algorithms were downloaded from Dr. Yasui's webpage (http://www.ualberta.ca/~yyasui/homepage.html). Weighted co-expression networks were constructed using WGCNA package [40], feature selection in modified SAM-GSR algorithm was implemented using penalizedSVM package [41].

# Results and Conclusions

**Real data**
*On the MSigDB C2 and C5 categories*
The study schema is presented in Figure 2. Upon gene sets in the MSigDB c2 category, we applied both SAM-GSR and modified SAM-GSR to the MS data. The selected pathways and genes by both algorithms are tabulated in Table 1. To evaluate both algorithms, we computed their predictive statistics on the training and test sets. As shown by the resulting statistics in Table 2A, modified SAM-GSR algorithm was superior to SAM-GSR algorithm in both the training set and the test set.

Additionally, the parsimony of modified SAM-GSR algorithm suffered. We conjectured that this might result from the automatic determination of reduced subset size in modified SAM-GSR algorithm. In contrast, one explicit criterion chosen by us herein for reduced set determination in SAM-GSR algorithm is parsimony. Moreover, the redundancy of information due to highly correlated genes in pathways may be attributable to the large size of reduced subset by modified SAM-GSR algorithm. On the contrary, SAM-GSR has more efficient control over this size by expelling genes with small values of individual SAM statistics outside.

*On weighted co-expression networks*
The curated pathways in major databases such as KEGG and GO tend to be enriched in the most prevalent-studied disease－cancers. This might put MS, a less investigated disease, in an inferior position. Especially when the goal is to incorporate the biological information involved in gene sets to improve upon predictive performance of a classifier, absence of such information might deteriorate the performance of a pathway-based feature selection algorithm. To evaluate if the deficiency of relevant pathways in the publicly available databases for MS has any impact on a pathway-based feature selection algorithm, we used weighted co-expression networks to cluster all genes annotated by HGU133plus2 package into 791 mutually exclusive subsets. Then, upon those gene subsets we reanalyzed the MS data using both SAM-GSR algorithms. The results are also presented in Table 2A.

Interestingly, there was no overlap between two subsets identified by these two algorithms, and the modified SAM-GSR showed a comparable predictive performance upon both training and test data to that using c2 and c5 categories. As illustrated in Figure 3, 30 genes intersected with genes involving in MSigDB c2 category, and 13 were overlapped with genes in c5 category. However, those genes were not selected in the analyses using c2 and c5 categories, where the pathway information is incomplete and biased towards cancers.

Comparing with the performance of other teams, we remark that if we had submitted the results of modified SAM-GSR analysis to sbv IMPROVER challenge, we would have been among top five. To conclude, complete pathway information is highly desired since the absence of meaningful biological information might result in declined discriminating capacity of a pathway-based feature selection algorithm. Provided such complete pathway information is unavailable currently, we suggest a statistical method is used for constructing statistical clusters, aiming at complementing those well-known biological pathways to achieve better performance.

*Comparison with other MS diagnosis signatures*
In this section, we compared several MS diagnosis signatures in the literatures with the ones we obtained using SAM-GSR algorithms. Here, we emphatically compared the performance of different signatures on the test set. The performance of those signatures by others were presented in Table 2B. Most relevantly, Guo et al. [42] obtained an 8-gene signature using the same training set. Then upon this 8-gene signature, its performance on the test set was evaluated. Overall, It ranked as the second worst among those different signatures by outperforming our original submission to sbv IMPROVER challenge. In this submission, a feature selection algorithm called Threshold Gradient Descent Regularization (TGDR) [43] and an integrated data including 6 microarray studies were used. To evaluate if different training data may have influence on the performance of a classifier, we reran TGDR using the data of E-MTAB-69 as the training set. The predictive performance improved dramatically based on the statistics in Table 2B. This partially justified that there is always data dependency for a classifier [44] and that batch effect among different experiments should be handled with caution when those data sets are combined [45].

**Synthesized data**

In Dinu et al [12], simulations were conducted to illustrate that SAM-GS algorithm outperforms GSEA in terms of identifying meaningful gene sets associated with the phenotype under study. Here since the focus was switched to explore the potential of SAM-GSR algorithms for feature selection, we conducted two simulations to explore if SAM-GSR algorithms can indeed distinguish true predictors apart from the false ones. Actual expression values of MS training data were used in the simulations, with an extra standardization to make the expression values of an individual gene have a mean of zero and a standard deviation of one.

In the first simulation, we chose the two genes selected by Tarca et al [39], i.e., F13A1 and GSTM1. These two genes are annotated within MSigDB c2 category, c5 category and hgu133plus2.db package. Then we randomly selected 998 genes from MS training data, those genes served as noises. The logit function for MS patients with controls as the baseline was given as following,

$$f_{MSvsControl} = 1.28 X_{F13A1} - 1.2 X_{GSTM1}$$

the values of coefficients of F13A1 and GSTM1 were simulated using an uniform distributed random variable in the range of -3 to 3. Furthermore, both genes are involved in many gene sets in both c2 (>40) and c5 (>20) categories.

In the second simulation, we chose two genes that are only involved in one or two gene sets in c2 and c5 categories as the relevant genes. The logit function for MS patients with controls as the baseline was given as following,

$$f_{MSvsControl} = -2.82 X_{RP9} - 2.41 X_{COX4I2}$$

the values before RP9 and COX4I2 were simulated using an uniform distributed random variable in the range of -3 to 3, again.

The results are tabulated in Table 3. An important finding here is when the true marker appears in many gene sets, SAM-GSR algorithm can easily identify it while modified SAM-GSR might miss it with high probability. Oppositely, when the true marker is rarely involved in several gene sets both SAM-GSR algorithms are highly likely to miss it. This explains why when the mutually exclusive gene sets constructed by co-expression network method were used, both SAM-GSR algorithms failed to identify at least one of two considered genes.

Another finding is the size of selected gene subset by SAM-GSR algorithm is substantially smaller than that by modified one, typically in the first extreme case where the true markers are involved in many gene sets. This is in consistent with the results of MS application. We remark this is because when a gene is involved in many gene sets, its probability of being selected increases. Namely, the chance of this gene having a big enough SAM statistic in several gene sets is obviously bigger than in one gene set.

# Discussion

Using a real-world application of a MS microarray data, modified SAM-GSR algorithm on complete co-expression network with the aid of a statistical method established excellent predictive performance. Based on the results from two simulations, we found modified SAM-GSR algorithm has two obvious drawbacks compared to SAM-GSR. One is high likelihood of failing to identify true markers and the other is poor control over the resulting subset size. However, these drawbacks are extrusive in the cases when true marker appears in many gene sets. On the opposite condition when true marker appears in rare gene sets, SAM-GSR algorithm also is ineluctable from missing true markers.

Given the fact the complete pathway information is far from completeness, a statistical method capable of constructing biologically meaningful gene networks is in demand. Based on the simulations, we hypothesize that a more complicated method than weighted co-expression network algorithm, which only considers correlations among genes in essence, may complement those annotated gene sets in relevant databases to produce a better classifier. The basic requirement for such method is it must take interactions and interplay among genes into account.

# Tables

## Table 1. The selected pathways and genes in MS data

|  | Pathways by SAM-GS | Genes by SAM-GSR | Genes by modified SAM-GSR |
|---|---|---|---|
| C2 | R: base excision repair, resolution of AP sites via the multiple nucleotide patch replacement, processive synthesis on the lagging strand, pol switching, repair synthesis for gap filling by DNA pol in TC NER, unwinding of DNA, removal of the flap intermediate from the C strand, DNA strand elongation, CD28 dependent P13K AKT signaling<br>O: Okamoto liver cancer multi-centric occurrence down *<br><br>Note: the cutoff of Q-value is 0.05. | POLD4 POLD2 POLD1 NTHL1 DNA2 MLST8 AKT3 RICTOR POLE MCM3 MCM5 GINS2 MCM2 RPA3 CLEC10A TAPBP TRIM25 AKAP17A<br><br>Note: the cutoff of $C_k$ is 0.2<br><br>(N=288, n=18) ** | Note: Since there are 271 genes in this final model and the performance of this model is inferior to others, the resulting genes are not presented here.<br><br>(N=288, n=271 using SCAD as penalty and n=112 using LASSO as penalty) |
| C5 | Transcription factor TFIID complex, amino acid derivative biosynthetic process, transcription from RNA polymerase activity, DNA polymerase activity, DNA directed DNA polymerase activity<br><br>Note: the cutoff of p-value is 0.01, if based on Q-value<0.05, No gene set is significant. | **ASMTL POLR2K ZNF76 BRCA1 TAF1 TAF11 EDF1 POLD4**<br><br>Note: the cutoff of $C_k$ is 0.05.<br><br>(N=59, n=8) ** | **TAF1** TAF6 TAF5 TAF8 **TAF11** TAF10 TAF13 TAF12 **EDF1** TAF9 OAZ1 GATM ASMTL ETNK1 TGFB2 GTF3A POLR2L **POLR2K** SNAPC4 **ZNF76** SNAPC3 POLR3C **BRCA1** TROVE2 GTF3C4 PTGES3 POLI POLH POLE POLA1 **POLD4** POLE4 POLE2 POLE3 POLD1 TEP1 POLG2 POLQ TERT REV3L<br><br>(N=59, n=40) |
| CN[1] | NA ***<br><br>Note: all genes were classified into 731 mutually exclusive get sets with the minimum size of 10, SAM-GS selected 4 of them. | LGALS3 RCL1 MSLN GLUL IKBIP RP2<br><br>Note: the cutoff of $C_k$ is 0.2.<br><br>(N=95, n=6) ** | ACVR2B CHMP4A EIF3J-AS1 ETV7 FANCE HOXC5 LINC00482 MCM3AP-AS1 PDCD4 PRCC SLC39A13 TTC9C VPS26B CD1C COQ6 DNPH1 EDEM2 HLA-DOA MLST8 POLR2E PSENEN SPINT2 TMC6 ACTL10 HIST1H1A KRBA1 SPRYD3 UBOX5 CHID1 GID4 LNX1 SGK494 TAF15 TRDMT1<br><br>(N=95, n=34) |

Note: * R stands for the pathways in Reactome; O is for the ones manually curated from the literature on gene expression in MSigDB c2 category. ** N represents the total number of unique genes annotated by hgu133plus2.db package in the selected pathways. *** NA stands for Not Applicable. Moreover, gene symbols in purple are the genes indicated as directly related to MS by the GeneCards database. The overlapped gene symbols between SAM-GSR and modified SAM-GSR were highlighted. [1] CN abbreviates for the gene sets constructing by the weighted co-expression network algorithm.

## Table 2. The performance statistics of selected genes on MS training and test sets

|  | Training set | | | | Test set | | | |
|---|---|---|---|---|---|---|---|---|
|  | Error (%) | GBS | BCM | AUPR | Error (%) | GBS | BCM | AUPR |
| A. The performance of SAM-GSR and modified SAM-GSR | | | | | | | | |
| C2: SAM-GSR | 20.45 | 0.121 | 0.701 | 0.896 | 46.67 | 0.464 | 0.500 | 0.644 |
| M-SAM-GSR | 0 | 0.066 | 0.747 | 0.992 | 46.67 | **0.291** | 0.520 | 0.612 |
| Lasso as penalty | 0 | 0.083 | 0.719 | 0.992 | **33.33** | **0.207** | 0.564 | **0.776** |
| C5: SAM-GSR | 13.64 | 0.134 | 0.673 | 0.904 | 46.67 | 0.464 | 0.500 | 0.579 |
| M-SAM-GSR | 0 | 0.046 | 0.800 | 0.992 | 43.33 | 0.365 | 0.577 | **0.703** |
| CN: SAM-GSR | 11.36 | 0.138 | 0.669 | 0.858 | 46.67 | 0.439 | 0.506 | 0.608 |
| M-SAM-GSR | 0 | 0.013 | 0.985 | 0.992 | **33.33** | **0.292** | **0.659** | 0.647 |

| B. The performance of other relevant signatures | | | | | |
|---|---|---|---|---|---|
| Study (size) | Training data used | Error (%) | GBS | BCM | AUPR |
| Lauria (n>100) | E-MTAB-69 | -- | -- | 0.884 | 0.874 |
| Tarca (n=2) | GSE21942 (on Human Gene 1.0 ST) | -- | -- | 0.629 | 0.819 |
| Zhao (n=58) | 7 other data besides E-MTAB-69 | 30 | -- | 0.576 | 0.820 |
| (n=84) | | 35 | -- | 0.549 | 0.636 |
| Tian (n=28)[1] | 5 other data besides E-MTAB-69 | 68.33 | 0.546 | 0.345 | 0.362 |
| (n=38)[2] | E-MTAB-69 only | 38.33 | 0.290 | 0.559 | 0.593 |
| Guo (n=8) | E-MTAB-69 * (10-fold CV error=13.64 %) | 46.67 | 0.462 | 0.499 | 0.504 |

Note: C2 represents the analysis using the pathways in MSigDB c2 category; C5 represents the analysis using the pathways in MSigDB c5 category; CN represents the analysis using the gene sets constructed by the weighted co-expression network algorithm. M-SAM-GSR abbreviates for modified SAM-GSR algorithm. GBS: Generalized Brier Score; BCM: Belief Confusion Metric; AUPR: Area Under the Precision-Recall Curve. * The predictive statistics on the test set for Guo's study were calculated based on the 8-gene signature they provided in their article. [1]This was the original submission by us to sbv IMPROVER using TGDR algorithm, it was ranked around 30 among 54 participants. [2]This was the reanalysis we did on the training set using TGDR to evaluate how the use of different training sets affects the performance of an algorithm.

Table 3. The performance of SAM-GSR and modified SAM-GSR on simulated data

| | SAM-GSR | | | Modified SAM-GSR | | |
|---|---|---|---|---|---|---|
| A. Simulation 1 (q-value in SAM-GS <0.05) | | | | | | |
| | F13A1 | GSTM1 | Size | F13A1 | GSTM1 | Size |
| C2 | Yes | Yes | 13 | No | Yes | 35 |
| C5 | Yes | Yes | 14 | No | No | 42 |
| CN | Yes | No | 11 | No | No | 47 |
| B. Simulation 2 | | | | | | |
| | RP9 | COX4I2 | Size | RP9 | COX4I2 | Size |
| C2 | No | Yes | 3 | No | Yes | 3 |
| C5 | No | Yes | 2 | No | Yes | 4 |
| CN[1] | No | Yes | 4 | No | Yes | 36 |

Note: Yes represents the gene is selected by algorithms; No means the gene is not selected; Size refers to the size of the reduced gene subset. [1] here, p-value <0.1 instead of q-value<0.05 in SAM-GS was chosen because there was no gene sets selected if using q-value.

# Figures

**Figure 1. Graphical illustration of both SAM-GSR and modified SAM-GSR algorithms. A. SAM-GSR for feature selection. B. Modified SAM-GSR.**

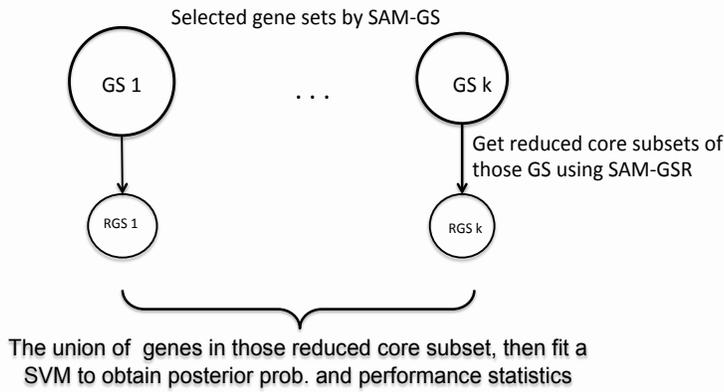
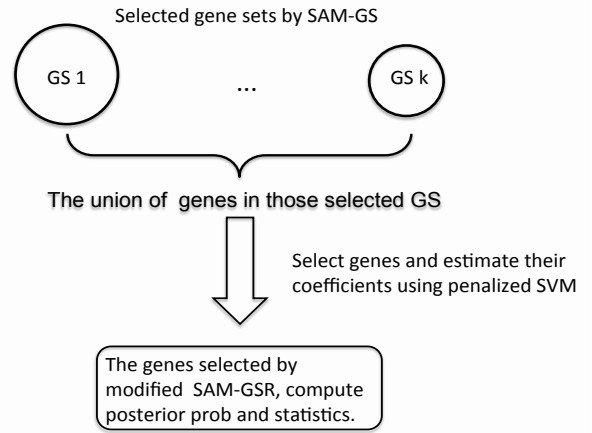

**Figure 2. Study schema illustrating how the analyses were conducted on the multiple sclerosis microarray data.**

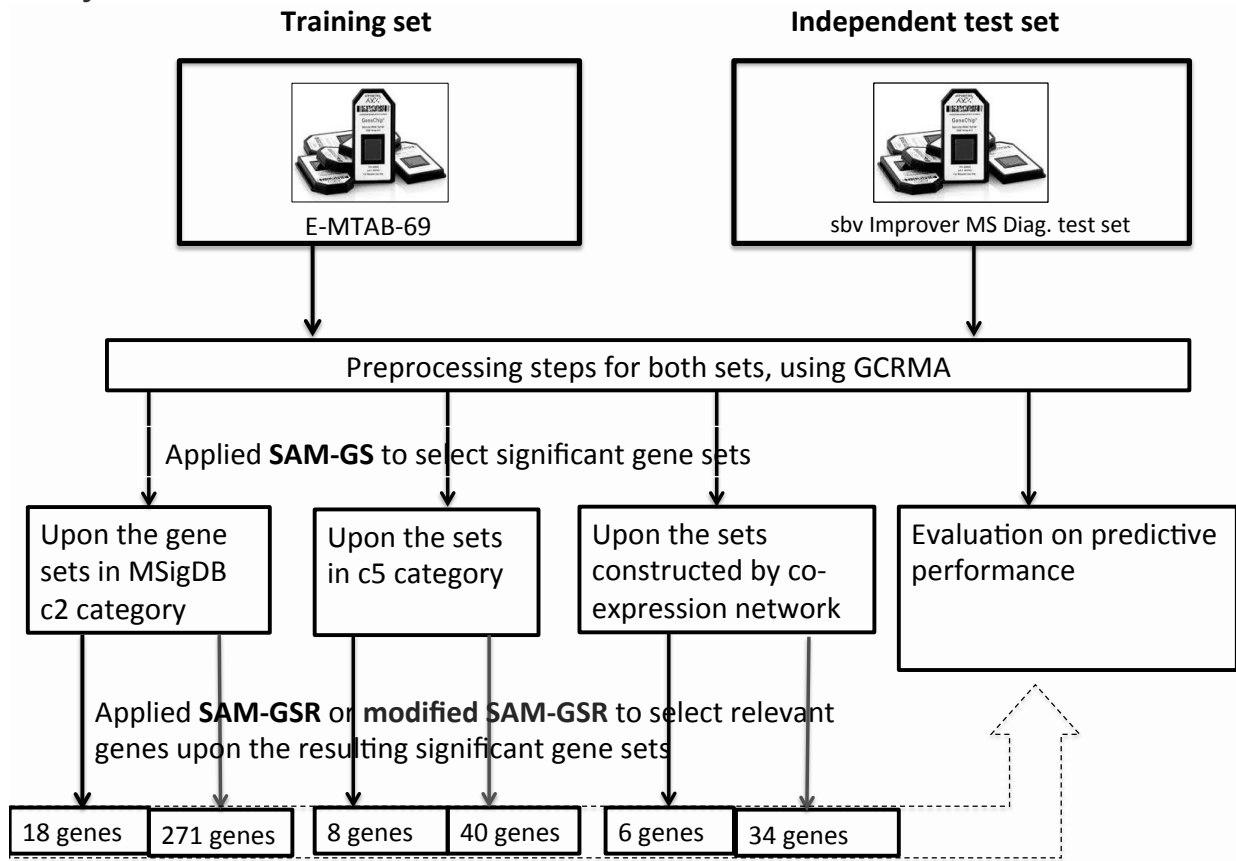

**Figure 3. Venn-diagram showing how the 34-gene signature selected by modified SAM-GSR algorithm intersected with genes annotated by MSigDB c2, c5 category and hgu133plus2 package.**

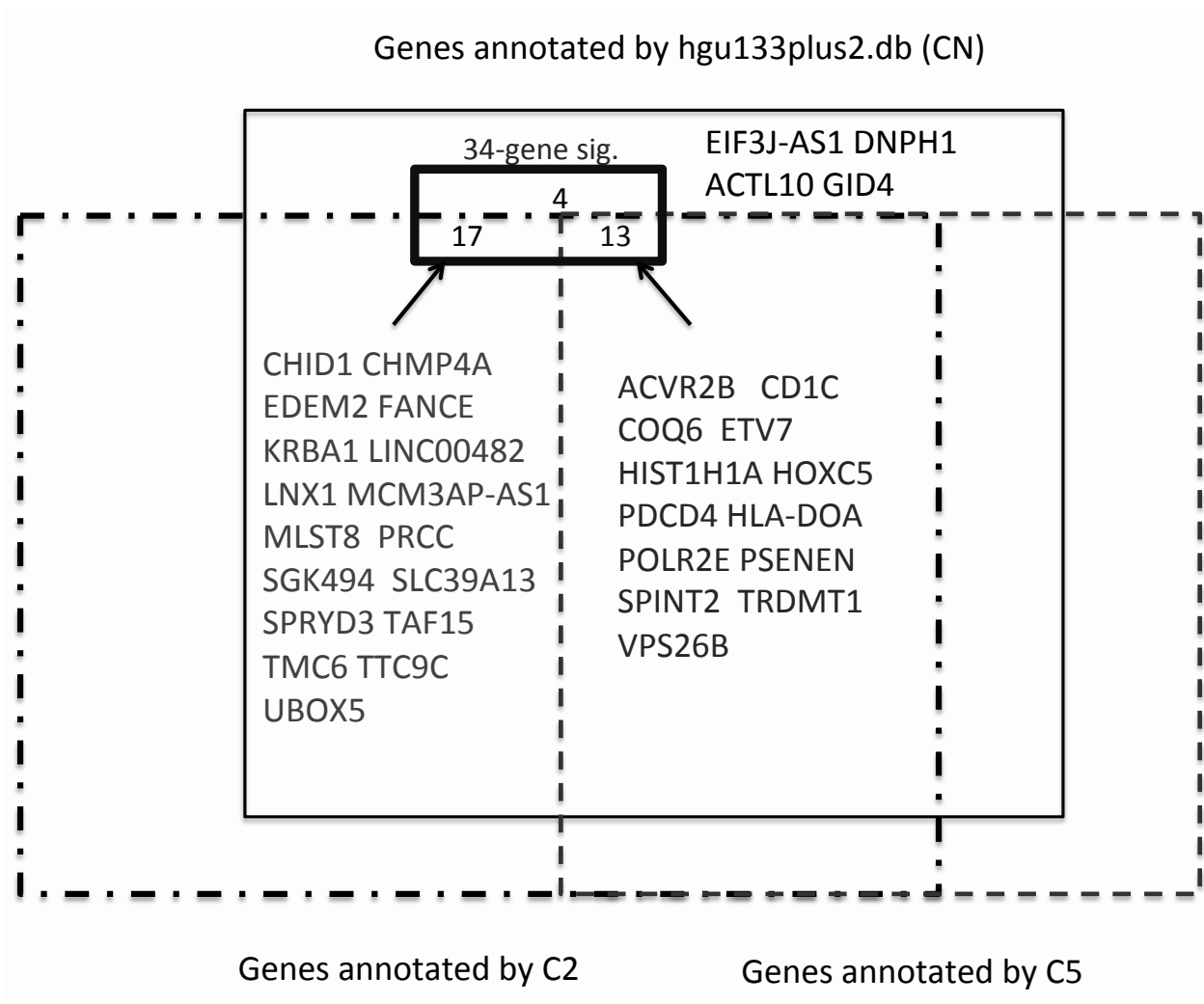